\input{aipcheck}

\documentclass[
    ,final            
  ]
  {aipproc}

\layoutstyle{8x11single}

\begin{document}

\title{Recent Results From The Daya Bay Experiment}


\author{Chao Zhang \\ (on behalf of the Daya Bay collaboration)}{
  address={Brookhaven National Laboratory, Upton, New York, USA}
}

\begin{abstract}
The Daya Bay reactor neutrino experiment has observed the disappearance of electron antineutrinos from nuclear reactors at $\sim$kilometer baselines. The relative measurement of the $\bar\nu_e$ rate and spectrum between near and far detectors allows for a precision measurement of the oscillation parameters $\sin^22\theta_{13}$ and $|\Delta{m}^2_{ee}|$. Two new antineutrino detectors (ADs) were installed in summer 2012, bringing the experiment to the final 8-AD configuration. With 621 days of data, Daya Bay has measured $\sin^22\theta_{13} = 0.084 \pm 0.005$ and $|\Delta{m}^2_{ee}| = 2.44^{+0.10}_{-0.11} \times 10^{-3}$ eV$^2$. 
This is the most precise measurement of $\sin^22\theta_{13}$ to date and the most precise measurement of $|\Delta{m}^2_{ee}|$ in this channel. 
Several other analyses are also performed, including an independent measurement of $\sin^22\theta_{13}$ using $\bar\nu_e$ samples tagged by neutron capture on hydrogen, a search for light sterile neutrinos, and a measurement of the absolute reactor antineutrino flux.
\end{abstract}

\maketitle


\section{Introduction}

The discovery of neutrino oscillation over the past several decades has established that neutrinos have mass and their flavors ($\nu_{e}, \nu_{\mu}, \nu_{\tau}$) mix. It represents one of the very few instances that indicate that the Standard Model is incomplete. The neutrino mixing can be characterized by three mixing angles ($\theta_{12}, \theta_{23}, \theta_{13}$), a possible CP violating phase $\delta_{CP}$, and two mass-squared differences $\Delta{m}_{21}$ and $\Delta{m}_{31}$. Determination of all their values with increasing accuracy was and continues to be the main goal of the experiments.

The value of $\theta_{13}$ has been a longstanding puzzle. Prior to 2012, only an upper limit of $\sin^22\theta_{13} < 0.15$ at 90\% C.L. was obtained by the CHOOZ~\cite{chooz} and the PALO VERDE~\cite{paloverde} experiment. The cleanest way to measure $\theta_{13}$ is through kilometer-baseline reactor neutrino oscillation experiments. The reactor $\bar\nu_e$ oscillation at $\sim$km is dominated by the $\theta_{13}$ terms, with the survival probability given by 
\begin{equation} \label{eqn:survival}
P_{ee} = 1 
- \sin^22\theta_{13}(\cos^2\theta_{12}\sin^2\Delta_{31} + \sin^2\theta_{12}\sin^2\Delta_{32})
-\cos^4\theta_{13}\sin^22\theta_{12}\sin^2\Delta_{21}
\end{equation}
where $\Delta_{ij}=1.267\Delta m^2_{ij}\frac{L(m)}{E(MeV)}$. Unlike the accelerator neutrino experiments, the reactor measurements are independent of $\theta_{23}$ and the CP-violating phase $\delta_{CP}$, and only slightly dependent on the neutrino mass hierarchy and the matter effect. 
In 2012, all three second-generation reactor $\theta_{13}$ experiments, Double Chooz, Daya Bay and RENO, reported clear evidences of $\bar\nu_{e}$ disappearance~\cite{dchooz,reno,dayabay} at $\sim$kilometer baselines with only a few month's running. In particular, Daya Bay excluded $\theta_{13}=0$ by 5.2 standard deviation with 55 days of data and provided the most precise measurement of $\theta_{13}$. In 2013, Daya Bay updated the results with 217 days of data, with a spectral analysis to measure the oscillation frequency, which led to the first direct measurement of the $\bar\nu_{e}$ mass-squared difference $|\Delta{m}^2_{ee}|$\footnote{The effective mass-squared difference $\Delta{m}^2_{ee}$ is defined such that $\sin^2\Delta_{ee} \equiv \cos^2\theta_{12}\sin^2\Delta_{31} + \sin^2\theta_{12}\sin^2\Delta_{32}$.}~\cite{dayabay2}. In this article, we report the new results from Daya Bay with 621 days of data and two newly installed antineutrino detectors. It represents four times more exposure than the previously reported results in Ref~\cite{dayabay2}.

\section{The Daya Bay Reactor Neutrino Experiment}

The Daya Bay reactor neutrino experiment was designed to provide the most precise measurement of $\theta_{13}$ among the existing and near future experiments, with sensitivity to $\sin^22\theta_{13} < 0.01$ at 90\% C.L. This is achieved through a careful site choice and detector design. A detailed description of the Daya Bay experiment can be found in Ref~\cite{dayabay_nima,dayabay_cpc}.

The Daya Bay experiment is located near the Daya Bay nuclear power plant (NPP) in southern China. The six reactors, two at Daya Bay and four at Ling Ao, provide a total of 17.6 GW thermal power, making it one of the largest in the world. Prior to Aug 2012, Daya Bay had six functionally identical antineutrino detectors (ADs), each with a 20-ton gadolinium-doped liquid scintillating (Gd-LS) target region. A near-far arrangement of the ADs in three experimental halls (EHs) is used. Two ADs were placed in EH1 at a distance of 364 m from the two Daya Bay reactor cores. One AD was place in EH2 at a distance of about 500 m from the four Ling Ao reactor cores. And three ADs were places in EH3 1912 m from the Daya Bay cores and 1540 m from the Ling Ao cores. The relative measurement of $\bar\nu_e$ rate and spectrum between near and far ADs allows for the cancellation of the relatively large uncertainty (2--5\%) in predicting the absolute reactor $\bar\nu_e$ flux. The functionally identical design of the ADs further removes the correlated detector systematics.

Each AD consists of a cylindrical, 5-m diameter stainless steel vessel (SSV) that houses two nested acrylic cylindrical vessels. A 3-m diameter inner acrylic vessel (IAV) holds 20 tons of Gd-LS as the detection target. It is surrounded by a region with 20 tons of liquid scintillator (LS) inside a 4-m diameter outer acrylic vessel (OAV). Between the SSV and OAV, 37 tons of mineral oil (MO) shields the LS and Gd-LS from external radioactivity. Each AD has 192 8-inch PMTs mounted on the side walls of the SSV. Optical reflectors are place at the top and bottom of the SSV, which effectively increases the photo-coverage to 12\%. On average, 160 photoelectrons are collected per MeV, which results in an energy resolution of $\sim$8\%$/\sqrt{E (\textrm{MeV})}$. Each AD has three automated calibration units (ACUs) mounted on top of the SSV lid. Weekly calibration is performed through remote deployment of an LED diffuser ball, a $^{68}$Ge source, and a combined source of $^{241}$Am$^{13}$C and $^{60}$Co into the Gd-LS and LS liquid volumes along three vertical axes. The position accuracy of the deployment is $<7$ mm~\cite{dayabay_acu}.

The ADs in each EH are submerged in a water pool, with at least 2.5 m of high-purity water in all directions to shield against ambient radiation. Each water pool is optically separated with Tyvek sheets into inner and outer water shields (IWS and OWS) and instrumented with PMTs to function as Cherenkov-radiation detectors. The detection efficiency for long-track muons is $>99.7$\%~\cite{dayabay_muon}. The water pool is covered with an array of resistive plate chambers (RPC) for independent muon tagging.

Antineutrinos are detected via the inverse beta decay (IBD) interaction, $\bar\nu_e + p \to e^+ + n$. The coincidence of the prompt signal from the $e^+$ and the delayed signal from neutron capture on Gd provides a distinctive $\bar\nu_e$ signature. It is crucial to characterize the detector response between ADs due to the relative measurement strategy of the Daya Bay experiment. Besides regular calibration with ACU sources ($^{68}$Ge, $^{60}$Co, $^{241}$Am$^{13}$C), special calibration sources ($^{137}$Cs, $^{54}$Mn, $^{40}$K, $^{241}$Am$^{9}$Be and Pu-$^{13}$C) were used in summer 2012 with AD1 and AD2 in EH1. Furthermore, a manual calibration system (MCS) was installed on AD1 during the special calibration period. The MCS allows for a full-volume deployment of sources to study the position dependence of the energy reconstruction. In addition, gamma and alpha peaks which could be identified in data ($^{40}$K, $^{208}$Tl, n capture on H, C, and Fe, $^{212}$Po, $^{214}$Po, $^{215}$Po) are included. The difference in reconstructed energy between ADs is less than 0.2\% in the energy range of reactor $\bar\nu_e$.

Interpretation of the observed prompt energy spectra requires precise characterization of the absolute energy scale for $e^+$, $e^-$, and $\gamma$. 
There are two major sources of nonlinearity between the true energy of the particle and the reconstructed energy: scintillator nonlinearity and electronics nonlinearity. The scintillator nonlinearity is particle and energy dependent, and is related to intrinsic scintillator quenching and Cherenkov light emission. It is modeled based on Birks' formula and constrained by a combined fit with mono-energetic gamma peaks and the $^{12}$B $\beta$-decay spectrum, where the $^{12}$B is a spallation product after muon interaction with carbon. The electronics nonlinearity is introduced due to the interaction of the scintillation light time profile and the charge collection of the front-end electronics. It is modeled with Monte Carlo and with the measurements from special single-channel FADC runs. The combined energy nonlinearity is cross-validated with internal radioactive $^{214}$Bi, $^{212}$Bi and $^{212}$Tl $\beta$-decay spectra, Michel electron spectrum in the ADs, and standalone bench-top Compton scattering measurements. The absolute energy scale uncertainty, which is correlated among all ADs, is constrained to be less than 1\% in the majority of the energy range of reactor $\bar\nu_e$.

Daya Bay took data in the aforementioned 6-AD configuration from December 2011 to July 2012. All previous results~\cite{dayabay,dayabay2,dayabay_cpc} were based on the data in the 6-AD period. In summer 2012, two additional ADs were installed in EH2 and EH3, which completed the final 8-AD configuration of the experiment. The data taking resumed after October 2012. The new results are based on the complete data set of the 6-AD period with the addition of the 8-AD period from October 2012 to November 2013, a total of 621 days.

\section{The Neutrino Oscillation Analysis}

IBD candidates are selected with the following criteria. First, events caused by PMT light emission are efficiently rejected. Candidates are then selected by requiring a prompt-like signal (0.7--12 MeV) in coincidence with a delayed-like signal (6--12 MeV) separated by 1--200 $\mu$s. Candidates are removed if the delayed-like signal occurs (i) within 0.6 ms from a water pool muon, (ii) within 1 ms from an AD muon ($E_{\mu}>20$ MeV), or (iii) within 1 s from an AD showering muon ($E_{\mu}>2.5$ GeV). Finally, a multiplicity cut is performed to select only isolated candidate pairs. The combined efficiency in selecting $\bar\nu_e$ Gd-capture events is $80.6\% \pm 2.1\%$. The relative uncertainty in the efficiency between ADs is $0.2\%$, which is the dominant factor in determining oscillation parameters. A total of 1.1 million (150k) IBD candidates are selected in the near (far) halls, representing the largest $\bar\nu_e$ sample among all previous and current reactor neutrino experiments. The time-dependent signal rates are strongly correlated with reactor $\bar\nu_e$ flux, which is predicted from the thermal power, fuel burn-up, and exchange and enrichment records provided by the NPP.

Daya Bay is a low-background experiment. The largest background comes from accidental coincidence of singles events, which accounts for 2.3\% (1.4\%) of the candidates in the far (near) hall. The accidental background, however, can be statistically calculated with high precision. A second background comes from neutrons from the $\sim$0.7 Hz Am-C calibration sources inside the ACUs on top of the ADs. They can occasionally mimic IBD events by inelastically scattering with nuclei in the shielding material and then capturing on Fe/Cr/Mn/Ni, producing two $\gamma$-rays that both enter the scintillating region. This background, which used to be the second largest in Daya Bay, is largely reduced after summer 2012 when the Am-C sources were removed from the two off-central-axis ACUs for the ADs in the far hall. It accounts for 0.2\% (0.03\%) of the far (near) hall candidates. Cosmogenic background in Day Bay is low due to good overburden of each underground hall (EH1: 250 m.w.e; EH2: 265 m.w.e; EH3: 860 m.w.e). The background from beta delayed-neutron emitters $^{9}$Li and $^{8}$He is 0.4\% (0.4\%) in the far (near) hall ADs. The fast neutrons produced by untagged muons account for 0.1\% (0.1\%) of the far (near) hall  candidates. Finally, the background due to $(\alpha, n)$ nuclear reaction is 0.1\% (0.01\%) in the far (near) hall ADs.

\begin{figure}[htb] \label{fig:oscillation}
  \centering
  \includegraphics[width=\textwidth]{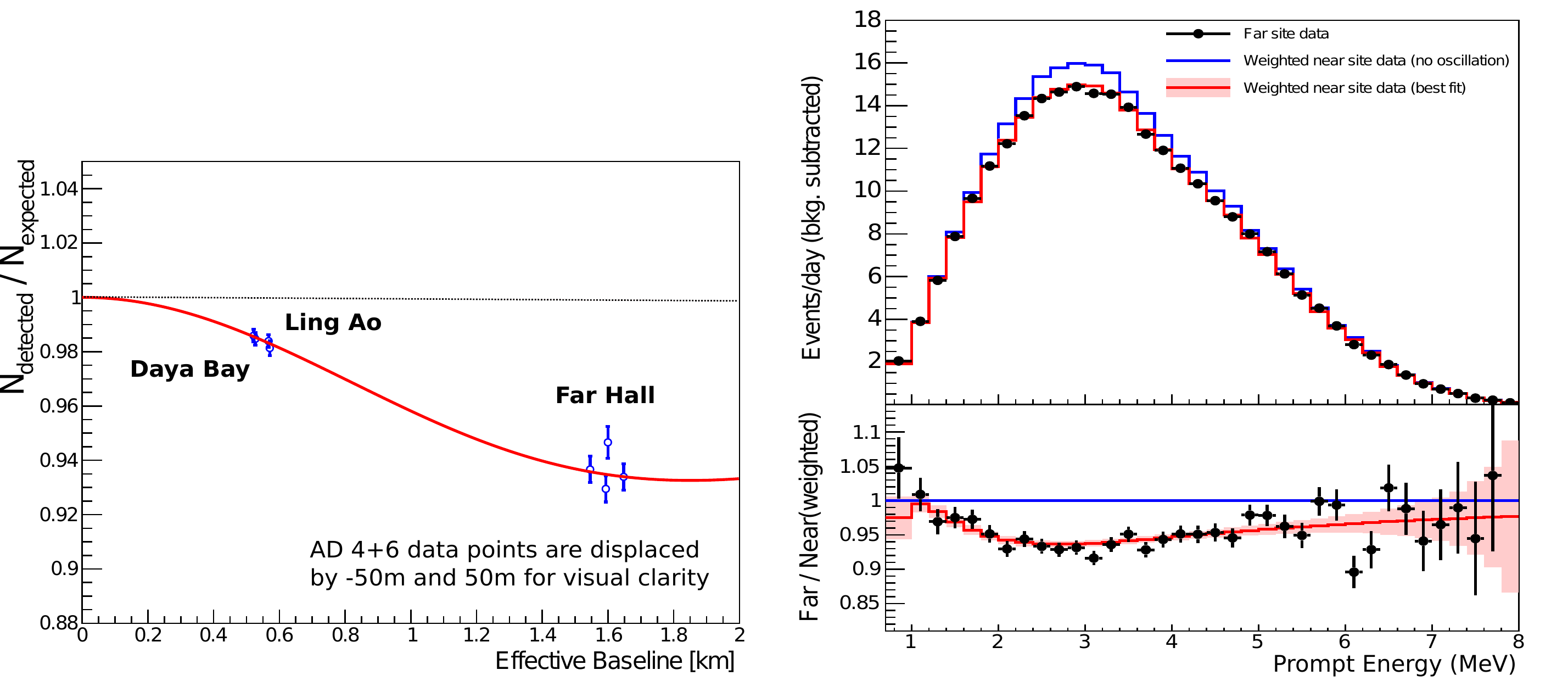}

  \caption{(left) Ratio of the detected to expected $\bar\nu_{e}$ signals at the 8 antineutrino detectors (ADs) located in three experimental halls as a function of effective baseline. The expected signal is corrected with the best-fit reactor $\bar\nu_{e}$ flux normalization.  The oscillation survival probability at the best-fit value is given by the red curve.
  (right) The top panel shows the measured background-subtracted spectrum at the far site compared to the expected spectrum based on the near site data both without oscillation and with the best-fit oscillation included. The bottom panel shows the ratio of the far site spectrum to the weighted near site spectrum.  The red curve shows the expectation at the best-fit oscillation values from the rate and spectral analysis.}
\end{figure}

The left plot of Figure~\ref{fig:oscillation} shows the ratio of the detected to expected $\bar\nu_{e}$ signals at the 8 ADs located in the three EHs, as a function of effective baseline. Consistent signal rates are measured for ADs in the same EH. The signal rate at the far site shows a clear $\sim$6\% deficit with respect to the near sites. The right plot of Figure~\ref{fig:oscillation} shows the measured background-subtracted spectrum at the far site compared to the expected spectrum based on the near site data both without oscillation and with the best-fit oscillation included. In the bottom panel, the ratio of the far site spectrum to the weighted near site spectrum is shown. The rate deficit and spectrum distortion are highly consistent with the oscillation interpretation.

\begin{figure}[htb] \label{fig:contour}
  \centering
  \includegraphics[width=0.7\textwidth]{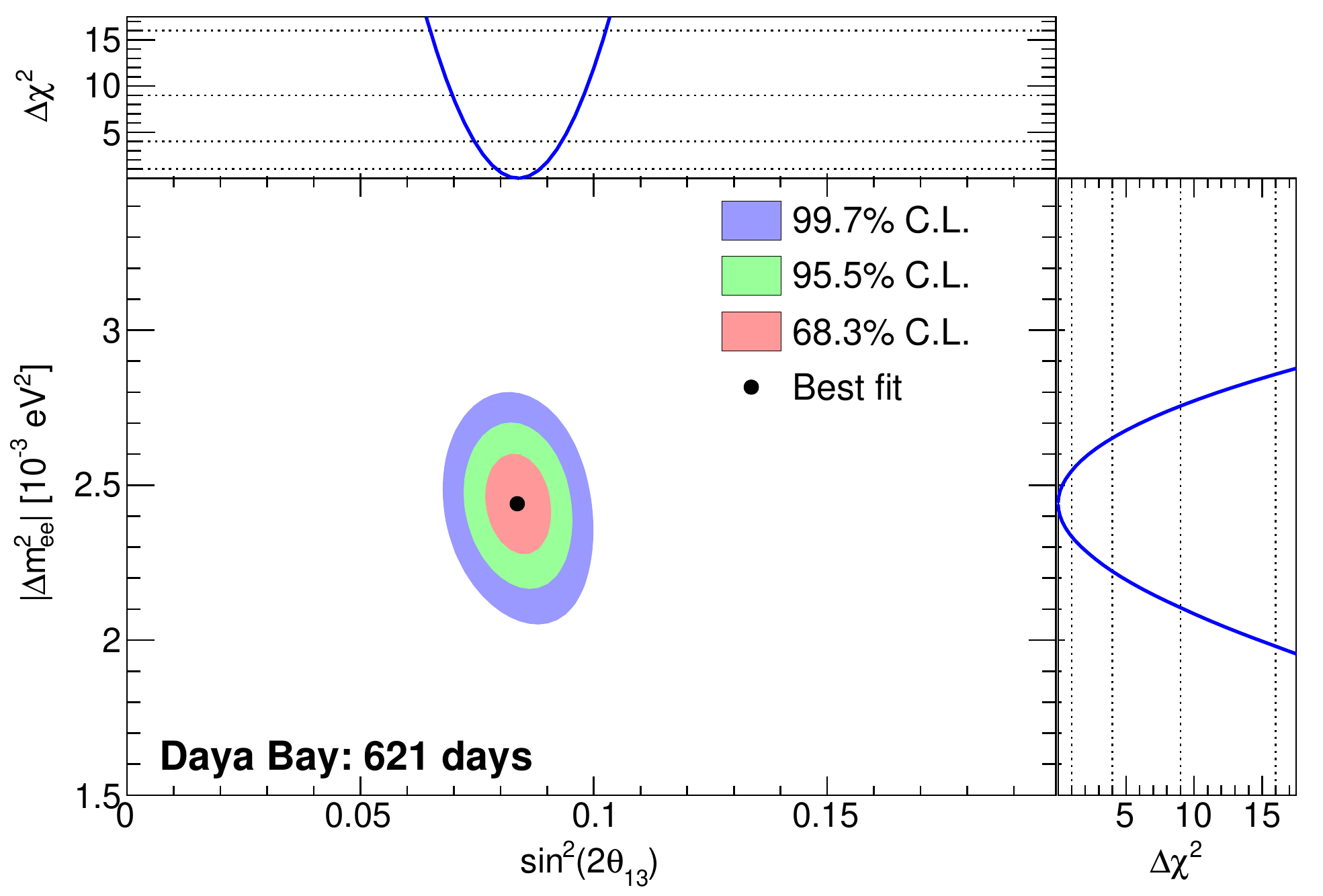}

  \caption{Allowed regions for $\sin^22\theta_{13}$ and $|\Delta{m}^2_{ee}|$ at the 68.3\%, 95.5\%, and 99.7\% confidence level, obtained from comparison of the rates and prompt energy spectra measured by the near and far site antineutrino detectors. The best estimate of the oscillation parameters is given by the black dot. The adjoining panels show the dependence of $\Delta\chi^2$ on each of the parameters.}
\end{figure}

The oscillation parameters are extracted from a purely relative comparison between data at the near and far halls. The observed prompt energy spectra of the near halls are extrapolated to the far hall and compared with observation. This process is done independently for each prompt energy bin. A covariance matrix, generated from a large Monte Carlo data set incorporating both statistical and systematic variations, is used to account for all uncertainties. An alternative analysis using nuisance-parameter-based $\chi^2$ method yields consistent results. Figure~\ref{fig:contour} shows the allowed regions for $\sin^22\theta_{13}$ and $|\Delta{m}^2_{ee}|$ at the 68.3\%, 95.5\%, and 99.7\% confidence level. The best fit values are $\sin^22\theta_{13} = 0.084 \pm 0.005$ and $|\Delta{m}^2_{ee}| = 2.44^{+0.10}_{-0.11} \times 10^{-3}$ eV$^2$ with $\chi^2/\textrm{NDF} = 134.7/146$. Although last known, the precision in the $\theta_{13}$ measurement (6\%) is now the best among all three mixing angles. The $|\Delta{m}^2_{ee}|$ measurement is highly consistent with and of comparable precision to the muon neutrino disappearance experiments. Under the assumption of normal (inverted) neutrino mass hierarchy, this result is equivalent to $|\Delta{m}^2_{32}| = 2.39^{+0.10}_{-0.11} \times 10^{-3}$ eV$^2$ ($|\Delta{m}^2_{32}| = 2.49^{+0.10}_{-0.11} \times 10^{-3}$ eV$^2$).

An independent measurement of $\sin^22\theta_{13}$ is performed using the $\bar\nu_e$ IBD samples tagged by neutron capture on hydrogen (nH). Inside the Gd-LS region, $\sim$15\% of neutrons capture on H instead of Gd. In addition, the full 20-ton LS region is used as detection target. Due to the high accidental background caused by the long neutron capture time ($\sim$200 $\mu$s) and low $\gamma$-ray energy (2.2 MeV) after n capture, the prompt candidate is required to have $E > 1.5$ MeV and the distance between the prompt and delayed candidates is required to be $<0.5$ m. The analysis uses the complete 217-day data set of the 6-AD period. A clear rate deficit is observed at the far hall. The rate-only analysis yields $\sin^22\theta_{13} = 0.083 \pm 0.018$, which is highly consistent with the results from nGd analysis. Details of the nH analysis can be found in Ref.~\cite{dayabay_nH}.

If a fourth (sterile) neutrino exists, it could cause additional spectral distortion betweens the ADs at different baselines. An analysis to search for such sterile neutrinos is performed using a minimal extension of the standard model: the 3(active) + 1(sterile) neutrino mixing model. Since Daya Bay has multiple baselines, the search depends only on the relative spectral difference between three sites and is largely independent of the reactor related uncertainties. The analysis uses the complete 217-day data set of the 6-AD period. The relative spectral distortion due to the disappearance of $\bar\nu_e$ is found to be consistent with that of the three-flavor oscillation model. The exclusion contours for $\sin^22\theta_{14}$ and $|\Delta{m}^2_{41}|$ are determined using both the Feldman-Cousins method and the CLs method~\cite{CLs}. The derived limits cover the $10^{−3}$ eV$^2$ < $|\Delta{m}^2_{41}|$ < 0.3 eV$^2$ region, which was previously largely unexplored. Details of the sterile neutrino analysis can be found in Ref.~\cite{dayabay_sterile}.

\section{Measurement of the Absolute Reactor Antineutrino Flux}

\begin{figure}[htb] \label{fig:dayabay_flux}
  \centering
  \includegraphics[width=0.7\textwidth]{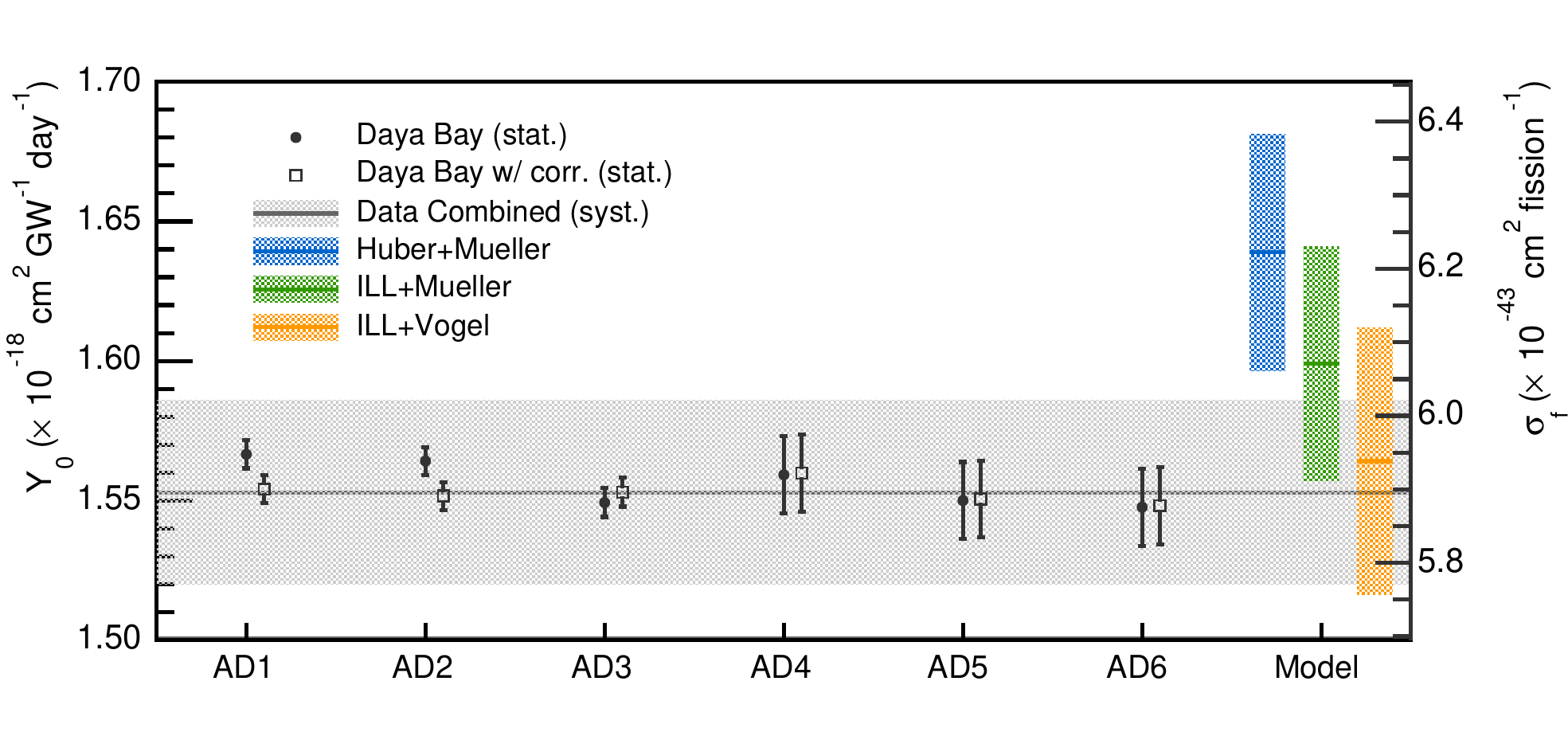}

  \caption{The measured reactor $\bar\nu_e$ event rate at each AD after correcting for the $\bar\nu_e$ survival probability, re-expressed as Y$_0$ (cm$^2$ GW$^{-1}$ day$^{-1}$) and $\sigma_f$ (cm$^2$ fission$^{-1}$). The solid and open circles show the data without and with correction for the difference in the effective fission fractions observed by each AD. The uncertainty of the measurement is shown as the gray band. Three theoretical model predictions are shown as a reference.}
\end{figure}

The large reactor $\bar\nu_e$ sample collected at Daya Bay allows for a precise measurement of the absolute reactor antineutrino flux. The analysis uses the complete 217-day data set of the 6-AD period. A total of 300k (40k) candidates are detected at the near (far) halls. Figure~\ref{fig:dayabay_flux} shows the measured reactor  $\bar\nu_e$ event rate at each AD after correcting for the $\bar\nu_e$ survival probability, re-expressed as Y$_0$ (cm$^2$ GW$^{-1}$ day$^{-1}$) and $\sigma_f$ (cm$^2$ fission$^{-1}$). The measurement among ADs is consistent within statistical fluctuations after correcting for the difference in the effective fission fractions. The uncertainty (2.3\%) of the measurement is dominated by the uncertainty in detection efficiency (2.1\%), which is correlated among all ADs. The measurement yields an average Y$_0$ = $1.553 \times 10^{-18}$ cm$^2$ GW$^{-1}$ day$^{-1}$ and $\sigma_f = 5.934 \times 10^{-43}$ cm$^2$ fission$^{-1}$, with the average fission fractions $^{235}$U : $^{238}$U : $^{239}$Pu : $^{241}$Pu = 0.586 : 0.076 :0.288 : 0.050.

Three theoretical model predictions are shown in Figure~\ref{fig:dayabay_flux} as a reference. The Huber~\cite{Huber} and ILL~\cite{ILL1, ILL2} models predict the $\bar\nu_e$ spectra for $^{235}$U, $^{239}$Pu and $^{241}$Pu, while the Mueller~\cite{Mueller} and Vogel~\cite{Vogel} models predict for $^{238}$U. The uncertainty in the model predictions is $\sim$2.7\%. The ratio (R) of the Daya Bay measurement to the Huber+Muller model prediction is $R = 0.947 \pm 0.022$, while $R = 0.992 \pm 0.023$ when compared to the ILL+Vogel model prediction.

\begin{figure}[htb] \label{fig:global_flux}
  \centering
  \includegraphics[width=0.7\textwidth]{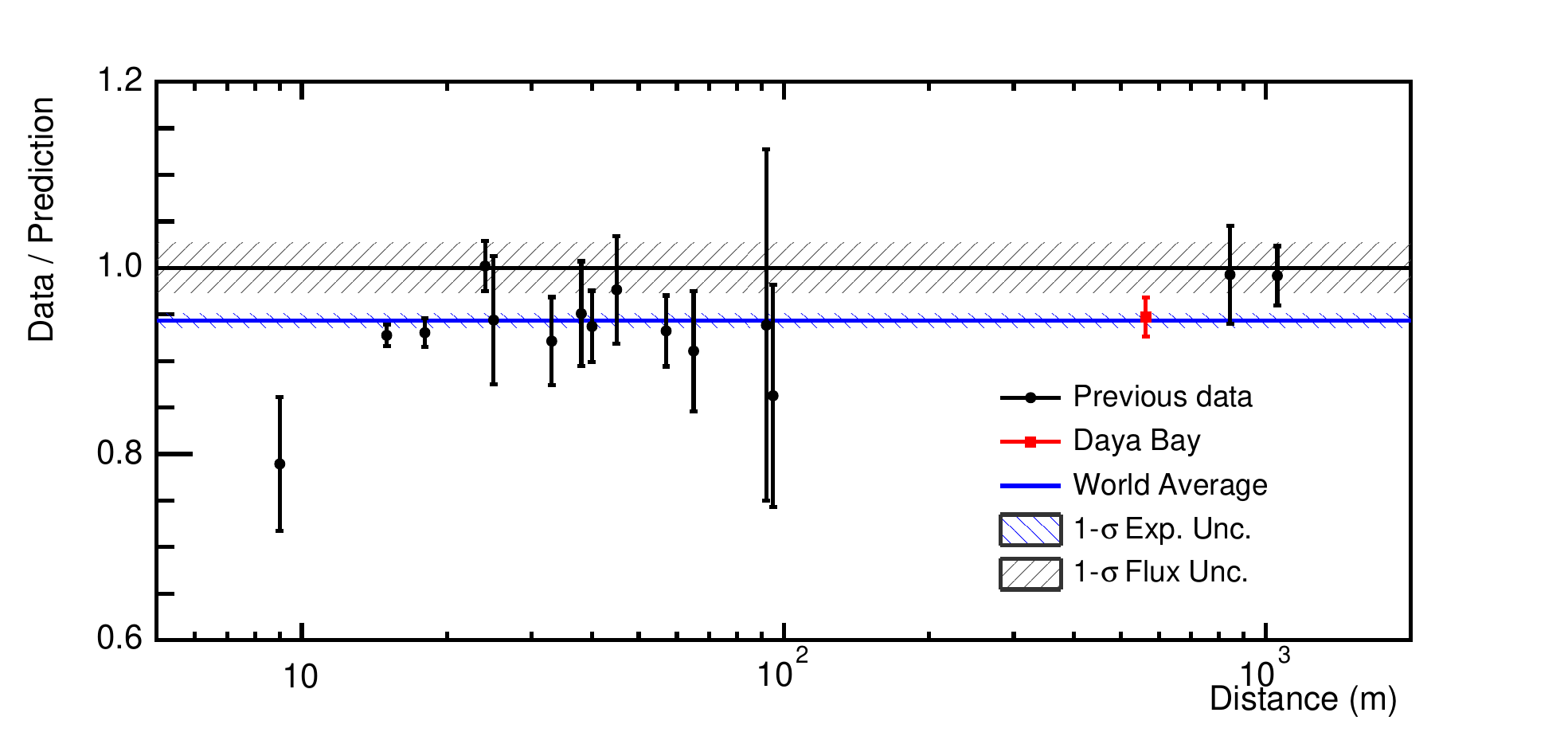}

  \caption{The reactor $\bar\nu_e$ interaction rate of the 21 previous short-baseline experiments~\cite{Mention, zhang} as a function of the distance from the reactor, normalized to the Huber+Mueller model prediction.~\cite{Huber,Mueller}.  Experiments at the same baseline are combined together for clarity. The Daya Bay experiment is placed at the effective baseline of 573 m. The rate is corrected by the $\bar\nu_e$ survival probability at the distance of each experiment, assuming standard three-neutrino oscillation. The horizontal bar (blue) represents the global average and its $1\sigma$ uncertainty. The 2.7\% reactor flux uncertainty is shown as a band around unity. }
\end{figure}

A comparison between the Daya Bay result and the 21 past reactor neutrino flux measurements is performed following the methods in Ref.~\cite{Mention, zhang}.
The result of the comparison is shown in Figure~\ref{fig:global_flux}. 
The Huber+Mueller model is used as the common reference model for all experiments. A neutron lifetime of 880.1 s~\cite{pdg} is used. 
The $\bar\nu_e$ survival probability is calculated with $\sin^22\theta_{13} = 0.089 \pm 0.009$ determined from the rate-only analysis using the complete 6-AD data set~\cite{dayabay2}.
The global average of the 21 past measurements with respect to the Huber+Mueller model prediction is determined to be $R = 0.943 \pm 0.008$ (experimental uncertainty), which is consistent with $R_{DYB} = 0.947 \pm 0.022$ from the Daya Bay measurement. 

\section{Summary}

The Daya Bay experiment uses the relative measurement of the $\bar\nu_e$ rate and spectrum between near and far detectors to precisely measure the oscillation parameters $\sin^22\theta_{13}$ and $|\Delta{m}^2_{ee}|$. Two new antineutrino detectors (ADs) were installed in summer 2012, bringing the experiment to the final 8-AD configuration. With 621 days of datas, Daya Bay has measured $\sin^22\theta_{13} = 0.084 \pm 0.005$ and $|\Delta{m}^2_{ee}| = 2.44^{+0.10}_{-0.11} \times 10^{-3}$ eV$^2$. This is the most precise measurement of $\sin^22\theta_{13}$ to date. The precision measurement of $\theta_{13}$ opens the door for future experiments to study neutrino mass hierarchy and leptonic CP violation. The $|\Delta{m}^2_{ee}|$ measurement is highly consistent with and of comparable precision to the muon neutrino disappearance experiments. By the end of 2017, Daya Bay expects to measure both $\sin^22\theta_{13}$ and $|\Delta{m}^2_{ee}|$ to precisions below 3\%.

Several other analyses are also performed. An independent analysis using the $\bar\nu_e$ IBD samples tagged by neutron capture on hydrogen has measured $\sin^22\theta_{13} = 0.083 \pm 0.018$. A sterile neutrino search has set stringent limits in the $10^{−3}$ eV$^2$ < $|\Delta{m}^2_{41}|$ < 0.3 eV$^2$ region. Finally, the absolute reactor antineutrino flux measurement has yielded consistent results with previous short-baseline reactor neutrino experiments.



\begin{theacknowledgments}
Daya Bay is supported in part by the Ministry of Science and Technology of China, the U.S. Department of Energy, the Chinese Academy of Sciences, the National Natural Science Foundation of China, the Guangdong provincial government, the Shenzhen municipal government, the China General Nuclear Power Group, Key Laboratory of Particle and Radiation Imaging (Tsinghua University), the Ministry of Education, Key Laboratory of Particle Physics and Particle Irradiation (Shandong University), the Ministry of Education, Shanghai Laboratory for Particle Physics and Cosmology, the Research Grants Council of the Hong Kong Special Administrative Region of China, the University Development Fund of The University of Hong Kong, the MOE program for Research of Excellence at National Taiwan University, National Chiao-Tung University, and NSC fund support from Taiwan, the U.S. National Science Foundation, the Alfred P. Sloan Foundation, the Ministry of Education, Youth, and Sports of the Czech Republic, the Joint Institute of Nuclear Research in Dubna, Russia, the CNFC-RFBR joint research program, the National Commission of Scientific and Technological Research of Chile, and the Tsinghua University Initiative Scientific Research Program. We acknowledge Yellow River Engineering Consulting Co., Ltd., and China Railway 15th Bureau Group Co., Ltd., for building the underground laboratory. We are grateful for the ongoing cooperation from the China General Nuclear Power Group and China Light and Power Company.
\end{theacknowledgments}






\hyphenation{Post-Script Sprin-ger}

\end{document}